\documentclass[universe,article,accept,moreauthors]{Definitions/mdpi}

\usepackage{graphicx}
\usepackage{subcaption}
\usepackage{comment}
\firstpage{1} 
\pubvolume{1}
\issuenum{1}
\articlenumber{0}
\pubyear{2026}
\copyrightyear{2026}
\datereceived{ } 
\daterevised{ } 
\dateaccepted{ } 
\datepublished{ } 

\Title{Probing dipole and quadrupole anisotropy in Gamma-ray bursts  from Swift  dataset}

\Author{Vedant Mokal $^{1}$\orcidA{}, Shantanu Desai $^{2}$*\orcidB{}}

\AuthorNames{Vedant Mokal, Shantanu Desai}

\address{%
$^{1,2}$ \quad Department of Physics, IIT Hyderabad, Kandi, Telangana-502284, India; ep23btech11034@iith.ac.in\\}

\corres{Correspondence: shntn05@gmail.com}

\abstract{Testing the validity of the cosmological principle's assumption of large-scale isotropy remains crucial for modern cosmology. We investigate the angular distributions of gamma-ray bursts using the GRB catalog from Neil Gehrels Swift Observatory (Swift) for  an independent probe of isotropy. Using the  HEALPix spherical harmonic decomposition, we estimate the dipole and quadrupole amplitudes and compare them against the null hypothesis obtained from 500 isotropic Monte Carlo realizations. Our results show 2.9$\sigma$ dipole and 7.2$\sigma$ quadrupole amplitude when applied to the raw data. To account for observational biases, we then create an exposure map using the pointing history, roll angle, and the partial coding fraction of the Swift Telescope. Reevaluating the null hypothesis using this map reduces the significance of these anisotropies to less than $1\sigma$. Therefore, our findings confirm statistical isotropy of the GRB sky using the Swift data, consistent with previous studies. We have also made the Swift  exposure map publicly available.}


\begin{document}


\section{Introduction}

Gamma-Ray bursts (GRBs) are brief, intense flashes of gamma radiation, which have been detected over a broad energy range from keV to TeV~\citep{zhang2018physics}. These GRBs can be divided into two categories, short and long, based on their duration, the time in which 90\% of the photon counts are received\citep{kouveliotou1993identification}.  Long-duration GRBs are linked to core-collapse supernova\citep{woosley2006supernova}, while short
GRBs originate from binary neutron star mergers\citep{nakar2007short}. However, there are a large number of exceptions to this dichotomy~\cite{Yang24}.

The distribution of gamma-ray bursts (GRBs) across the sky provides an independent way to test the cosmological principle. This principle assumes that the Universe is uniform and isotropic in all directions on large enough scales. It forms the basis for the standard Friedmann-Lemaître-Robertson-Walker (FLRW) metric\citep{weinberg2008cosmology}.

While the cosmic microwave background (CMB)~\citep{penzias1965measurement,smoot1992structure} and other tracers like radio sources~\citep{blake2002velocity}, Planck-selected galaxy clusters (using the Sunyaev-Zeldovich effect)~\cite{Bengaly17} and large-scale galaxy surveys~\citep{pandey2017testing,sarkar2019testing,franco2024probing} strongly support isotropy, new tensions have emerged based on analysis of recent data. These include quasar dipoles~\citep{secrest2021test,secrest2022challenge,kothari2024study} and large cosmic voids~\citep{keenan2013evidence,haslbauer2020kbc}. These tensions encourage ongoing careful testing with a large number of observational probes. (See Ref.~\cite{Barua26} and references therein for a recent summary of these anomalies).

GRBs are well-suited for this job. They can be detected over great distances and provide almost all-sky samples that are mostly free from standard redshift survey selection effects\citep{ghirlanda2006gamma}. Early BATSE analyses confirmed broad isotropy~\citep{Meegan1992, Briggs1996}. However, later studies found possible anisotropies in short-duration GRBs and unusual clustering patterns~\citep{Vavrek2008, Tarnopolski2017, Balazs2018, Horvath2024}. However, careful analysis suggested that some of these anomalies might result from instrumental biases rather than cosmic signals~\citep{Andrade2019}. In addition to the number distribution of GRBs, some studies  have also tested the isotropy of  GRB observables  such as duration, fluence,  and peak fluxes in various energy bands, and found them to be consistent with isotropy~\cite{Ripa17,Ripa19}. Nevertheless, some other works which have analyzed GRBs
continue to find deviations from isotropy and hemispherical asymmetry~\cite{Horvath26}.

Therefore, to investigate this in detail, a systematic search for dipole and quadrupole anisotropy was carried out using the GRB datasets of Fermi-GBM and BATSE in ~\cite{Mondal2026} (M26, hereafter). This work used spherical harmonic decomposition for their analysis and found that the dipole anisotropy amplitude was consistent with the null hypothesis within $1\sigma$. Although the quadrupole amplitude was found to be elevated with respect to the null distribution at $3.7\sigma$  and $3.0\sigma$, respectively, for the raw uncorrected BATSE and Fermi GBM data,  this quadrupole anisotropy vanished once corrections were made for the sky exposure. 

We now extend the analysis in M26  to the Swift GRB catalog described in \cite{lien2016third}. We measure the dipole and quadrupole amplitudes to detect any statistically significant asymmetries that might challenge the standard cosmological model.
Importantly, we address differences in the instrumental exposure, which can masquerade as real anisotropies. Since there was no publicly available exposure map, we also create a new effective exposure map for Swift based on its pointing history and the BAT detector's partial coding fraction.

The rest of this paper is structured as follows: Section~\ref{sec:dataset} covers the datasets, Section~\ref{sec:methodology} describes the statistical methods, Section~\ref{sec:results} presents the results of the multipole analysis and Section~\ref{sec:expcorrection} the impact of exposure corrections. Finally, section~\ref{sec:conclusions} summarizes our findings and explores their implications for the cosmological principle.

\section{The Swift GRB dataset}
\label{sec:dataset}
The Swift dataset is derived from observations by the \textit{Neil Gehrels Swift Observatory}, relying on detections from Burst Alert Telescope (BAT)~\cite{gehrels2004Swift,lien2016third}. Launched in November 2004 and remaining operational to the present day, Swift is uniquely designed for rapid automatic pointing and high-precision tracking, making its catalog very useful for multi-wavelength GRB studies. This continuously updated dataset provides a sample of bursts spanning roughly two decades of observation. For this analysis, we incorporate an extensive sample of 1759 GRBs recorded between December 17, 2004 and January 27, 2026. We do a holistic analysis taking into account all these  GRBs. In the Appendix, we present the analysis separately for short and long GRBs.

\begin{figure}[htbp]
    \centering

    \includegraphics[width=0.9\textwidth]{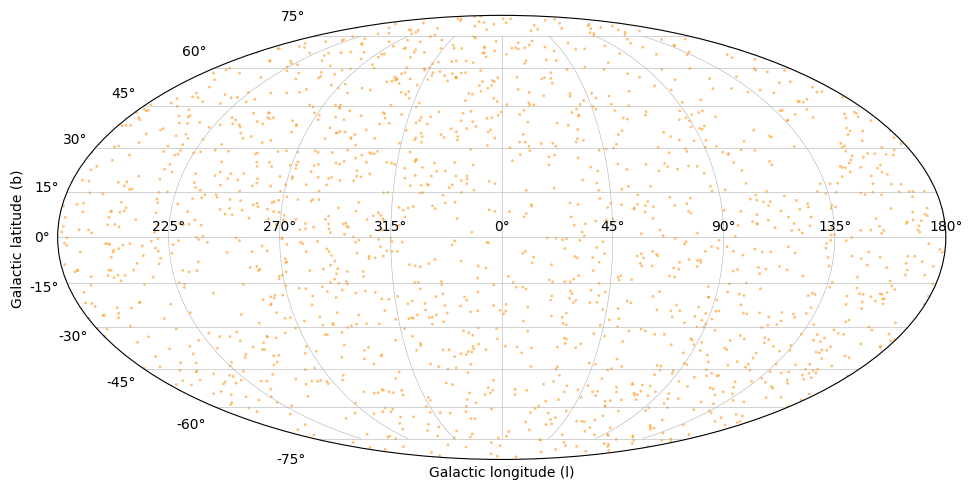}

    \vspace{0.5cm}

    \includegraphics[width=0.9\textwidth]{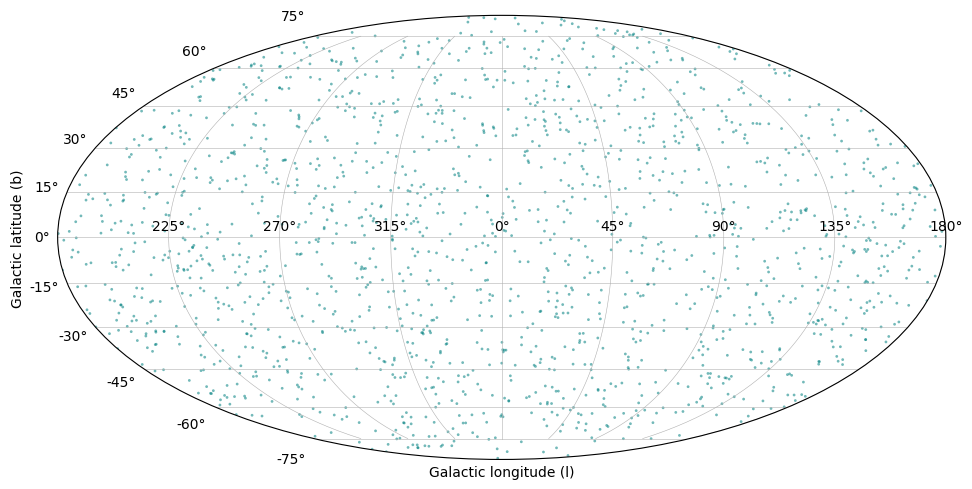}

    \vspace{0.5cm}

    \includegraphics[width=0.9\textwidth]{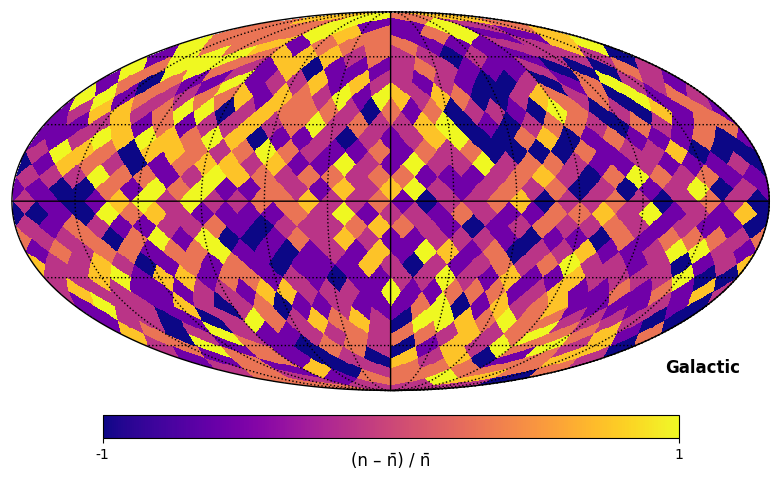}

    \caption{
    This figure shows the angular distribution of 1759 GRBs from the Swift GRB catalog (top panel) and a uniform random distribution of 1759 simulated points (middle panel), shown in Mollweide projection and Galactic coordinates. The bottom panel shows a HEALPix map of the fluctuation of GRBs from the Swift GRB catalog using $N_{\mathrm{side}} = 8$.
    }

    \label{fig:Swift_sky_distribution}
\end{figure}

\section{Methodology}
\label{sec:methodology}
For this work, we have applied the same prescription as that in M26 to the Swift dataset. We summarize the analysis procedure and defer to M26 for more details.

\subsection{Sky Pixelation and GRB distribution}

 We convert the GRB equatorial coordinates (RA, Dec) into Galactic coordinates ($l$, $b$), where $l$ ranges from $0^\circ$ to $360^\circ$ and $b$ ranges from $-90^\circ$ to $+90^\circ$. We used the HEALPix framework~\cite{Gorski2005} to split the sky into segments with equal area. We choose a resolution parameter $N_{\text{side}}$ dividing the sphere into $N_{\text{pix}} = 12 \times N_{\text{side}}^2$ pixels, with each pixel covering a solid angle of $\frac{41,253}{N_{\text{pix}}}$ square degrees. For our analysis we have chosen $N_{\text{side}}=8$, resulting in $N_{\text{pix}}=768$. In the Appendix we  test the robustness of our conclusions with different values of $N_{\text{side}}$.

Using the \texttt{ang2pix} function, we calculate the number of GRBs in each pixel. We then obtain the fluctuation of these counts compared to the average, creating a discretized distribution function $f(\theta, \phi)$, where $\theta$ ranges from $0^\circ$ to $180^\circ$ and $\phi$ ranges from $0^\circ$ to $360^\circ$. The fractional fluctuation for a given pixel at $(\theta, \phi)$ is defined as:
\begin{equation}
f(\theta, \phi) = \frac{n(\theta, \phi) - \bar{n}}{\bar{n}},
\label{eq:1}
\end{equation}
where $n(\theta, \phi)$ represents the count of the number in the direction of $(\theta, \phi)$ and $\bar{n}$ is the mean count across all pixels. Fig. \ref{fig:Swift_sky_distribution} shows the angular distribution of the 1759 GRBs in the  Swift catalog along with  an isotropic sky containing the same number of GRBs. Finally, we show the Hierarchical Equal Area isoLatitude Pixelization (HEALPix)~\cite{Gorski2005} map of $f(\theta, \phi)$ for the Swift GRB dataset using $N_{\mathrm{side}} = 8$.

 The Swift positional uncertainties (90\% error radii) in the dataset range from 0.46 to 10 arcminutes. The HEALPix grid we used ($N_{\mathrm{side}} = 8$), splits the  sky into 768 pixels. Each of these pixels subtends approximately 440 arcminutes on a side, and hence  the angular uncertainties have negligible effect on which pixel the GRB lies in, therefore we do not use the angular uncertainties in our analysis.

\subsection{Dipole and Quadrupole Estimation}
We then expand the  discretized distribution function, $f(\theta, \phi)$, into spherical harmonics, $Y_{\ell m}(\theta, \phi)$, stopping the expansion up to  the quadrupole  level ($\ell = 2$). This expansion can be expressed as:
\begin{equation}
f(\theta, \phi) = \sum_{\ell=0}^{2} \sum_{m=-\ell}^{\ell} a_{\ell m} Y_{\ell m}(\theta, \phi)
\end{equation}

In this representation, the monopole component is represented by $\ell = 0$, whereas the dipole component is represented by $\ell = 1$  and the quadrupole component by  $\ell = 2$. The expansion coefficients, $a_{\ell m}$, are found by integrating over the solid angle:
\begin{equation}
a_{\ell m} = \int_{0}^{2\pi} \int_{0}^{\pi} f(\theta, \phi) Y_{\ell m}^{*}(\theta, \phi) \sin\theta \, d\theta \, d\phi
\end{equation}

The overall dipole amplitude is then constructed from the dipole coefficients as:
\begin{equation}
A_{\text{dipole}} = \sqrt{|a_{1,-1}|^2 + |a_{1,0}|^2 + |a_{1,1}|^2}
\end{equation}

To compute this, we used the dipole estimator introduced by \citet{secrest2021test}. We do this with the \texttt{fit\_dipole} routine from the \texttt{healpy} Python package~\citep{Zonca2019}. This routine performs a linear least-squares regression to fit a monopole and dipole model directly to the HEALPix map.

Similarly, we  define the total quadrupole amplitude as the sum in quadrature of the five quadrupole coefficients:
\begin{equation}
A_{\text{quad}} = \sqrt{|a_{2,-2}|^2 + |a_{2,-1}|^2 + |a_{2,0}|^2 + |a_{2,1}|^2 + |a_{2,2}|^2}
\end{equation}
We computed the observed quadrupole amplitude ($A_{\text{quad}}^{\text{obs}}$) directly from the HEALPix fluctuation map via \texttt{healpy.anafast}, which computes the angular power spectrum ($C_\ell$) of a given HEALPix map.

\subsection{Hypothesis Testing using Monte Carlo simulations}
We test the significance of observed dipole and quadrupole by generating 500 Monte Carlo (MC) simulations of isotropically distributed skies for each dataset.  Each simulated sky included the same number of GRBs found in the actual observational catalogs: 1759 for Swift BAT.

For each simulated sky, we calculated the dipole ($A_{\text{dipole}}^{\text{sim}}$) and quadrupole ($A_{\text{quad}}^{\text{sim}}$) amplitudes using the spherical harmonic decomposition method previously described. We then combined these simulated amplitudes to create Probability Density Functions (PDFs). These PDFs show the expected distribution of dipole and quadrupole amplitudes, assuming cosmic isotropy. We smoothed  these PDFs using  Kernel Density Estimation (KDE)~\citep{silverman1986density}.

We compare the measurements from the observed GRB catalog to their corresponding isotropic PDFs. Our null hypothesis assumes that the spatial distribution of GRBs is isotropic and that any observed dipole or quadrupole moment is simply a result of random statistical fluctuations. This distribution of the null hypothesis obtained from the 500 synthetic datasets constructed.

To assess the significance, we calculate the  $p$-value for an observed amplitude ($x_{\text{obs}}$) by computing the integral:
\begin{equation}
p = P(X \ge x_{\text{obs}}) = \int_{x_{\text{obs}}}^{\infty} f(x) \, dx
\label{eq:p}
\end{equation}
where $f(x)$ is the expected distribution for the null hypothesis, as described later,  for most of the cases we calculate the $p$-value based on counting the total number of instances in which the dipole amplitude for the synthetic isotropic skies exceeds the observed value.


\section{Results}
\label{sec:results}
\begin{figure}[htbp] 
    \centering
      \includegraphics[width=0.85\linewidth]{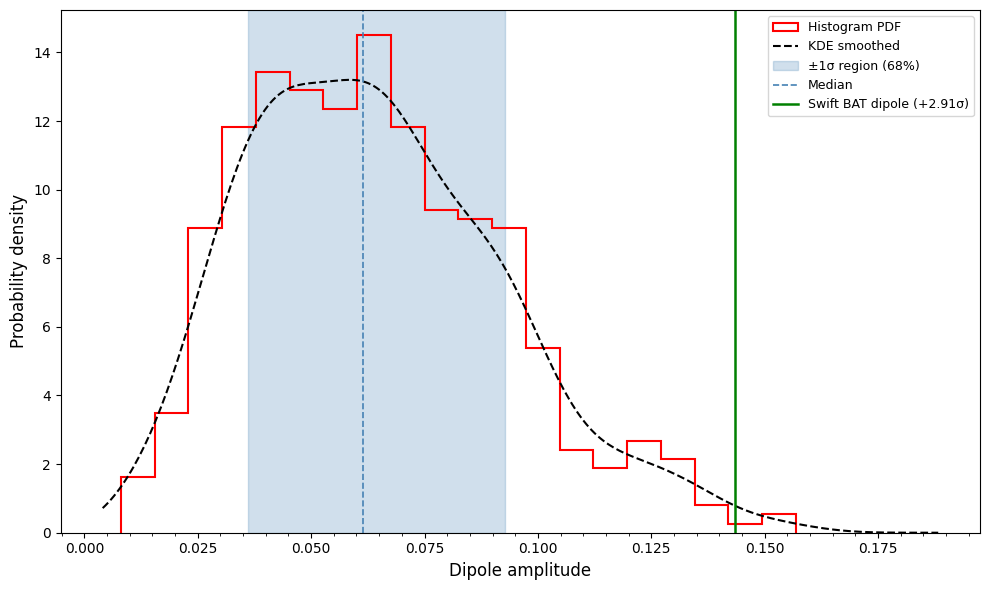}
    \caption{Observed Dipole vs PDF obtained from Monte Carlo simulations 0f 500 skies for Swift GRB DATA}
    \label{fig:Swift_dipole}
\end{figure}

\begin{figure}[htbp] 
    \centering

    \includegraphics[width=0.85\linewidth]{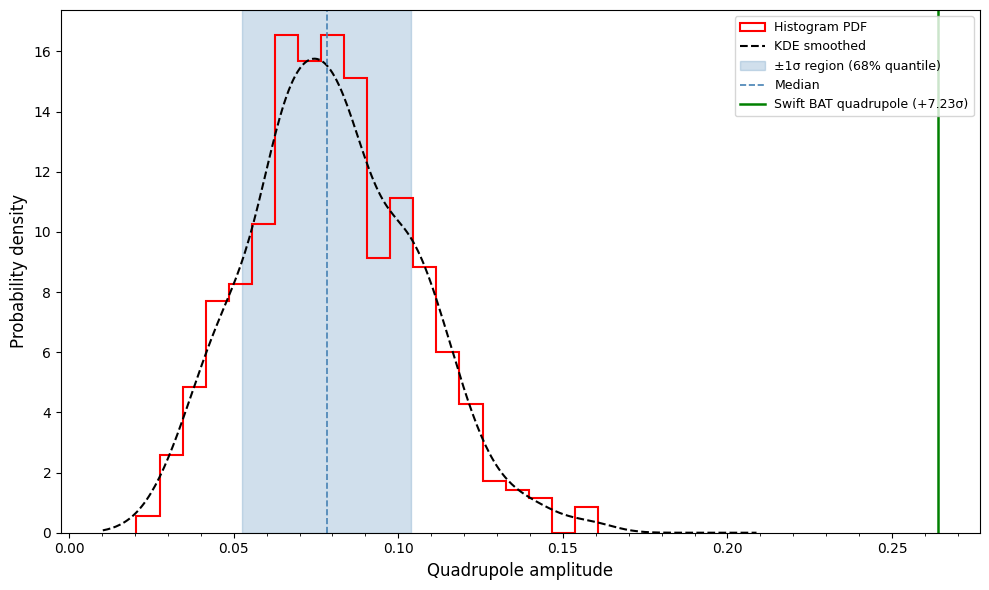}
    \caption{Observed quadrupole vs Gaussian fit to Monte Carlo simulations for 500 skies for Swift GRB DATA}
    \label{fig:Swift_quadrupole}
\end{figure}

We first calculate the dipole and quadrupole amplitudes for the 500 synthetic isotropic skies using the HEALPix built-in functions. Figs.~\ref{fig:Swift_dipole} and  ~\ref{fig:Swift_quadrupole} show the resulting PDFs of dipole and quadrupole amplitudes derived from the simulated isotropic skies, respectively. The step histogram in each of these figures depicts the binned distributions of amplitudes across these simulations. The dashed black curve shows the kernel density estimate (KDE), which provides a representation of the underlying PDF. 
The shaded blue region indicates a $1\sigma$ interval around the median value of the distributions. The  $1\sigma$ intervals have been computed from the 68\% quantile values of the reconstructed dipole/quadrupole for the isotropic skies.  We then apply the same procedure to the Swift dataset.  We find the dipole and quadrupole amplitudes for our mock isotropic skies to be $0.062 \pm 0.028$ and 
$0.079 \pm 0.026$ respectively.  We now compare these with the real data.
\subsection{Dipole}
We find the value of the  observed dipole amplitude to be  $0.143$ showing a $2.9\sigma$ deviation from the null hypothesis. The observed $p$-value is about 0.006. At first glance, this implies  that the dipole amplitude measured from Swift GRB dataset shows a marginal disagreement with a perfectly isotropic distribution.

\subsection{Quadrupole}
For the same dataset we find that the observed quadrupole amplitude to be equal to  $0.264$ showing a  $7.2\sigma$ deviation from perfect isotropy. As the detected quadrupole is at the very end of the tail, we cannot find the $p$-value using Eq.~\eqref{eq:p}, since none of the simulated skies yield a quadrupole amplitude greater than the observed value. So we fit a Gaussian to the simulated data for calculating the $p$-value. The $p$-value is equal to  $2.59 \times 10^{-13}$ corresponding to a significance of 7.2$\sigma$~\cite{Cowan11}.

This suggests the presence of a strong and non-negligible large-scale quadrupole anisotropy. 
However, when faced with a similar result using the raw data, M26 showed that once we take into account corrections due to non-uniform exposure, the significance  of the observed isotropy became negligible. Therefore, we must also apply a similar correction due to instrumental exposure.
However, unlike BATSE, there is no publicly available full-sky exposure map for Swift  to the best of our knowledge.~\footnote{Although a Swift exposure map based on the 157 month all sky X-ray survey appears in literature~\citep{Lien25}, this is not publicly available.}  Therefore, we create our own exposure map for Swift.

\section{Exposure Correction}
\label{sec:expcorrection}
To account for instrumental biases, we perform a detailed analysis using the exposure correction. We construct the exposure map by combining the telescope's pointing history, roll angle, and its Partial Coding Fraction (PCF). The Partial Coding Fraction is the fraction of the detector array that is illuminated by a source. Its data contains the fractional exposure as a function of how far the source is from the boresight, i.e., the center of the detector.

\subsection{Construction of the Exposure Map}

The exposure map is generated on a HEALPix grid ($N_{\rm side} = 8$) through the following procedure, which accounts for the instrument's asymmetric 2D response and the spacecraft's orientation:

\begin{enumerate}
    \item \textbf{2D PCF Coordinate Mapping:} We map the PCF image \footnote{Available at \url{https://Swift.gsfc.nasa.gov/proposals/bat_cal/index.html}} from pixel space to angular space using the FITS World Coordinate System (WCS) metadata \citep{greisen2002representations}. This establishes the exact coordinates on the PCF image of the pixel that is some angular distance away from boresight.
    
    \item \textbf{Pointing and Roll Integration:} For each valid point with positive exposure times ($t_{\rm exp}$), we extract the spacecraft roll angle ($\phi_{\rm roll}$). We find all HEALPix pixels within the $50^{\circ}$ field of view.
    
    \item \textbf{Detector Frame Transformation:} For each pixel $p$ within the FOV, we calculate the angular separation ($\theta_p$) and sky position angle ($PA_p$) relative to the telescope's boresight. We rotate these coordinates into the detector frame by subtracting the roll angle: $\theta_{\rm det} = PA_p - \phi_{\rm roll}$. We obtain $\Delta x$ and $\Delta y$ as the area where that pixel will fall on the PCF. These are obtained as follows~\cite{Mondal2026}:
    \[ \Delta x = \theta_p \sin(\theta_{\rm det}) \]
    \[ \Delta y = \theta_p \cos(\theta_{\rm det}) \]
    
    \item \textbf{Interpolation and Weighted Co-addition:} We interpolate the PCF array at ($\Delta x, \Delta y$) co-ordinates to assign pixel illumination weight $w_p \in [0,1]$. We then add the weighted exposure $t_{\rm exp} \times w_p$ back to the HEALpix sky pixel (in RA and DEC) to create the exposure map. 
\end{enumerate}

\begin{figure}[htbp]
    \centering
    \includegraphics[width=0.85\linewidth]{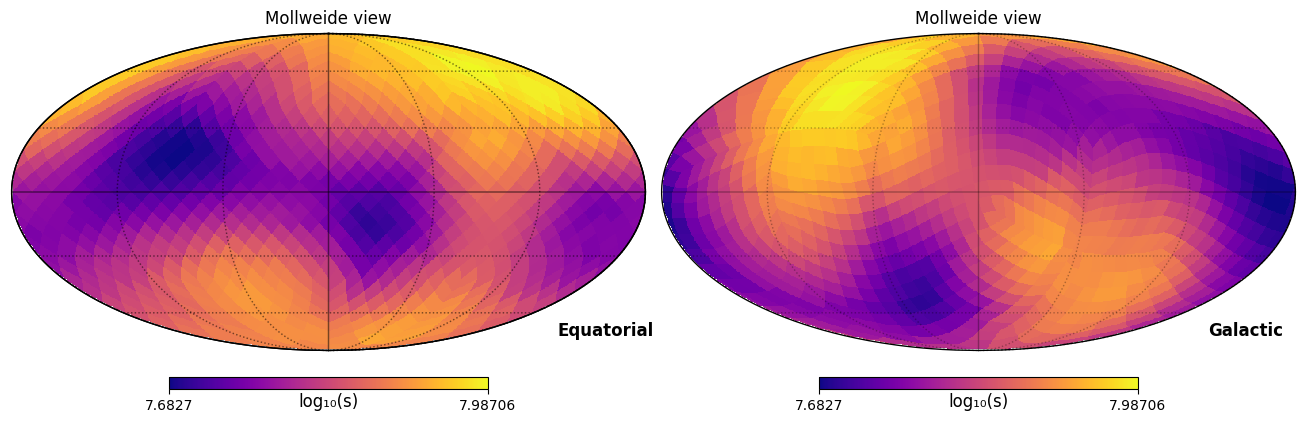}
    \caption{HEALPix probability map of the Swift sky exposure in equatorial coordinates, constructed by co-adding all pointings from the Swift pointing history weighted by the Partial Coding Fraction. The map reflects the accumulated PCF-weighted exposure across the mission lifetime and is used as the instrument response for the rejection-sampling procedure described in this section.}
    \label{fig:Swift_exposure_map}
\end{figure}

The crux of the above process is determining where, on the detector, all the pixels within the $50^{\circ}$
 field of view will be mapped.
 Once we find where they lie on the detector plane we assign them the weight. The resulting exposure map is shown in Fig. \ref{fig:Swift_exposure_map}, with units of [seconds $\times$ coding fraction], and serves as a measure of the detector bias towards observing certain areas in the sky.

It must be noted that our exposure map agrees qualitatively with the Adaptive Kernel Density Estimation (AKDE) completeness map of Swift BAT derived in \cite{Bagoly26}. Despite the different approaches in constructing these,  both the maps recover similar large scale structure. This agreement provides an independent validation that our PCF-weighted exposure map captures instrumental bias of the Swift/BAT detector.

We also note that the Swift telescope  has undergone significant degradation over the mission lifetime. \citet{Moss2022} (cf. Section 2.1) have pointed out  that the number of active CdZnTe detectors used in Swift/BAT declined from 32,768 to approximately 18,000 by 2020, reducing overall sensitivity. \citet{Lien25} (cf. Appendix 1) have also reported that the  radiation damage is causing additional coding damage. Our current exposure map does not account for these sources of  degradation and assumes a uniform sensitivity. A more time dependent approach incorporating this degradation is beyond the scope of this analysis.

\begin{figure}[htbp]
    \centering
    \includegraphics[width=0.85\linewidth]{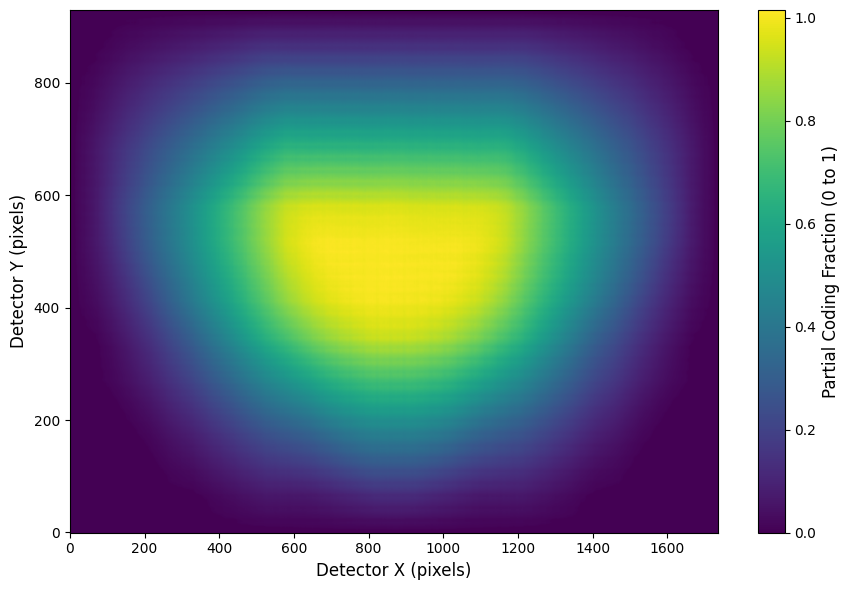}
    \caption{The Swift  2-D Partial Coding Fraction (PCF) map. Its asymmetric sensitivity profile demonstrates why spacecraft roll angle must be accounted for to project exposure onto the sky.}
    \label{fig:Swift_pcf_profile}
\end{figure}

\subsection{Exposure-Convolved Isotropic Reference Skies}
We now generated 500 mock isotropic skies convolved with the Swift exposure map using a 2D rejection-sampling technique similar to that used in M26:
\begin{enumerate}
    \item The trial positions $(\alpha, \delta)$ are drawn uniformly on the sphere to ensure an isotropic prior.
    \item The normalized exposure probability $P_{\rm exp}(p)$ is retrieved for the corresponding HEALPix pixel.
    \item Points are accepted if a random deviate $u' \sim \mathcal{U}(0,1) < P_{\rm exp}(p)$, ensuring the synthetic sky matches the true instrumental  sensitivity.
\end{enumerate}

We follow the same methodology as  described in Section~\ref{sec:methodology} to find the dipole and quadrupole amplitude for each simulated sky using the exposure corrected distribution. The observed Swift  amplitudes were then compared with the null hypothesis from these distributions to determine the significance  and  $p$-value.

\begin{figure}[htbp] 
    \centering      
    \includegraphics[width=0.85\linewidth]{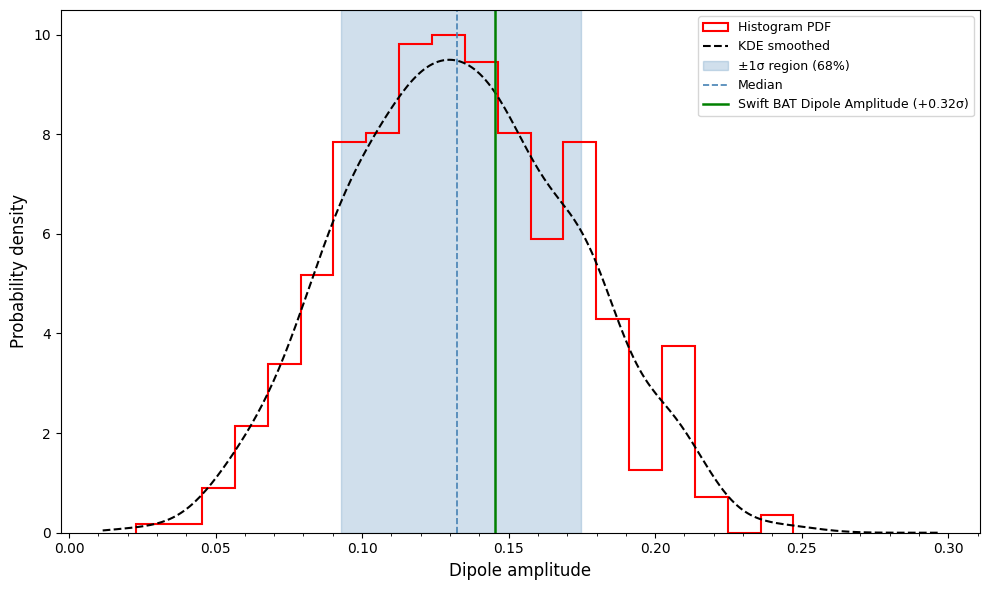}
    \caption{Exposure corrected PDF of dipole amplitude for Swift}
    \label{fig:corrected_Swift_d}
\end{figure}

Fig.~\ref{fig:corrected_Swift_d} represents the PDF of exposure-convolved isotropic simulations and the real dipole amplitude of Swift. The observed dipole amplitude is now equal to $0.14 \pm 0.04$. 
The significance of the observed dipole now decreases to only  $0.2\sigma$ and the $p$-value becomes $0.39$,  and  is therefore consistent with a perfectly isotropic distribution.

\begin{figure}[htbp] 
    \centering

    \includegraphics[width=0.85\linewidth]{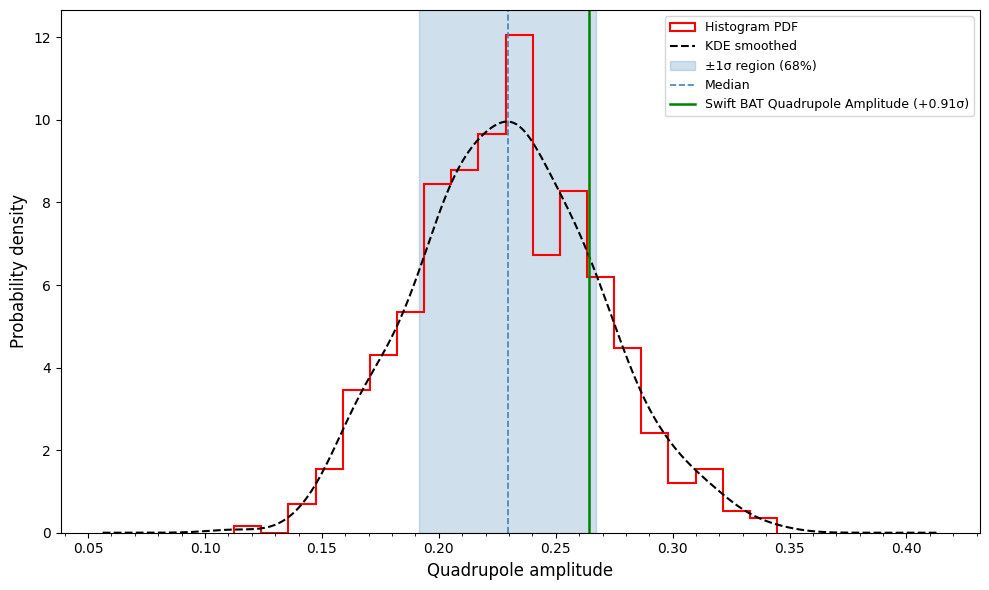}
    \caption{Exposure corrected PDF of Quadrupole amplitude for Swift}
    \label{fig:corrected_Swift_q}
\end{figure}

We build a PDF of the quadrupole amplitude the same way as that for the dipole. This PDF is shown in Fig. \ref{fig:corrected_Swift_q}, along with the observed quadrupole value for the real data. The quadrupole amplitude for the exposure corrected null distribution is now equal to $0.22 \pm 0.04$. 
The deviation  of the observed quadrupole with respect to the null distribution now reduces to only $1.0\sigma$ with a $p$-value of 0.178. Therefore, we find (similar to M26) that the statistically significant quadrupole anisotropy vanishes once exposure corrections are applied.

These results show that both the dipole and quadrupole anisotropies found in the uncorrected Swift GRB maps are artifacts due to uneven exposure of the Swift telescope. Once we carefully model this effect and include it in the isotropic simulations, the Swift GRB sky aligns with statistical isotropy. This highlights the important need to consider instrumental peculiarities in any analysis of large-scale anisotropy based on GRB datasets and agrees with the conclusions in M26. A tabular summary of our results can be found in Table~\ref{tab1}.


\begin{table}[htbp]
    \centering
    \caption{\label{tab1} Observed and Simulated Dipole and Quadrupole Amplitudes after comparing with exposure corrected distribution for the null hypothesis}
    \label{tab:anisotropy_results}
    \resizebox{\textwidth}{!}{%
    \begin{tabular}{llcccc}
        \toprule
        \textbf{Dataset} & \textbf{Multipole} & \textbf{Observed} & \textbf{MC median ($\pm$ 68\% quantile)} & \textbf{Dev.} & \textbf{$p$-value} \\
        \midrule
        \multirow{2}{*}{\textbf{Swift  (Raw)}} 
        & Dipole     & 0.143 & $0.062 \pm 0.028$ & $+2.9\sigma$ & 0.0060 \\
        & Quadrupole & 0.264 & $0.079 \pm 0.026$ & $+7.2\sigma$ & $2.59\times10^{-13}$ \\
        \midrule
        \multirow{2}{*}{\textbf{Swift  (Corrected)}} 
        & Dipole     & 0.143 & $0.13 \pm 0.04$ & $+0.2\sigma$ & 0.3880 \\
        & Quadrupole & 0.264 & $0.22 \pm 0.04$ & $+1.0\sigma$ & 0.1780 \\
        \bottomrule
    \end{tabular}%
    }
\end{table}

\section{Conclusions}
\label{sec:conclusions}
We studied the large-scale angular isotropy of gamma-ray bursts (GRBs) using the all-sky Swift  catalog (1759 GRBs) using the same prescription as M26. We measured the dipole and quadrupole amplitudes with a HEALPix-based framework. We also checked its significance against 500 Monte Carlo simulations of isotropic skies.

The raw data showed a dipole and quadrupole amplitudes with significance of $2.9\sigma$ and $7.2\sigma$, respectively. 
However, after creating a custom exposure map from the telescope's pointing history and Partial Coding Fraction, these significances dropped to $0.2\sigma$ and $1.0\sigma$, implying that there is no evidence for dipole or quadrupole anisotropy.
This indicates that the initial deviations using the raw data were due to the telescope's observation bias. This also agrees with the findings in M26, who showed that the quadrupole amplitude for BATSE GRB dataset is consistent with isotropic distribution, once we use the exposure corrected map.

These findings strongly reaffirm the statistical isotropy of the GRB sky for Swift selected GRBs. It shows no violation of the cosmological principle. This result shows an important theme in large-scale isotropy analyses. Apparent anisotropies in all-sky surveys can often be results of instrumental biases. Furthermore, our creation of a physical Swift  exposure map offers a framework for future studies of Swift dataset.

In the spirit of open science, we have made our Swift exposure map publicly available which can be found online at \url{https://github.com/Lithium-spirit/Swift-isotropy}

\vspace{6pt} 





\authorcontributions{Both authors contributed equally to this manuscript.}

\funding{This research received no external funding.}

\dataavailability{  The GRB dataset used for this analysis was downloaded from \url{https://swift.gsfc.nasa.gov/archive/grb_table/}. The Swift observatory pointing history, roll angle and partial coding fraction (PCF) calibration files used to generate the exposure maps are available via the NASA High Energy Astrophysics Science Archive Research Center (HEASARC). The resultant exposure map for Swift dataset has been made available  at \url{https://github.com/Lithium-spirit/Swift-isotropy}.}

\acknowledgments{
We thank the High Energy Astrophysics Science Active Research Center ( HEASARC)
for making Swift data publicly accessible. We are grateful to the anonymous referees for constructive feedback and several useful comments on our manuscript.
}

\conflictsofinterest{The authors declare no conflicts of interest.} 

\appendixtitles{Yes} 
\appendixstart
\appendix
\section{Stability test with different $N_{\text{side}}$  values}
The Healpix resolution parameter $N_{\text{side}}$ determines the number of pixels on the celestial sphere. A higher value gives a better angular sampling whereas a lower value gives a smoother map. Higher values might increase the statistical noise for smaller dataset while lower values can slightly affect the estimation of dipole and quadrupole. Therefore, testing the results with different $N_{\text{side}}$ values is important to ensure  that our findings are not influenced by the chosen resolution of $N_{\text{side}}=8$.

To evaluate this we repeat the entire process of dipole and quadrupole analyses using exposure convolved isotropic skies for three higher resolutions:  $N_{\text{side}}=16, 32, 64$.

\begin{figure}[htbp] 
    \centering

    \includegraphics[width=0.85\linewidth]{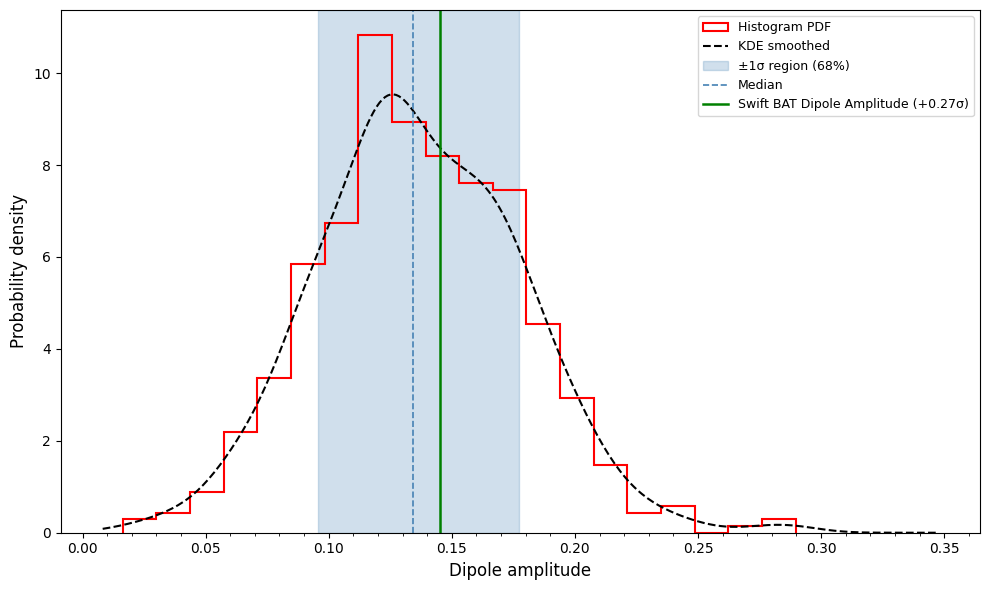}
   
    \caption{Observed dipole for  vs Exposure corrected PDF of dipole amplitude for Swift for $N_{\text{side}}=16$.}
    \label{fig:corrected_dipole_N16}
\end{figure}

\begin{figure}[htbp] 
    \centering

    \includegraphics[width=0.85\linewidth]{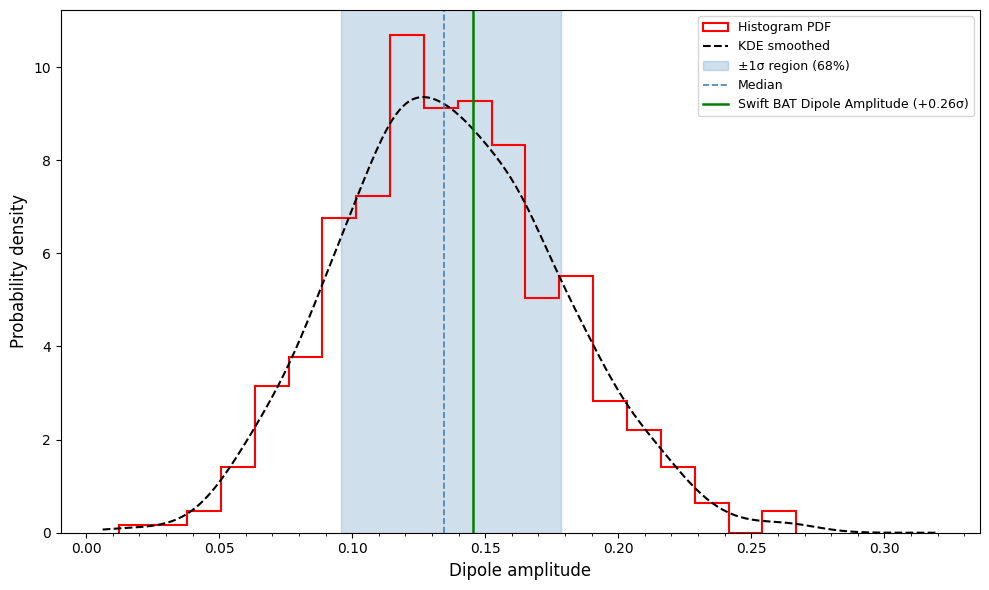}
   
    \caption{Observed dipole for  vs Exposure corrected PDF of dipole amplitude for Swift for $N_{\text{side}}=32$.}
    \label{fig:corrected_dipole_N32}
\end{figure}

\begin{figure}[htbp] 
    \centering

    \includegraphics[width=0.85\linewidth]{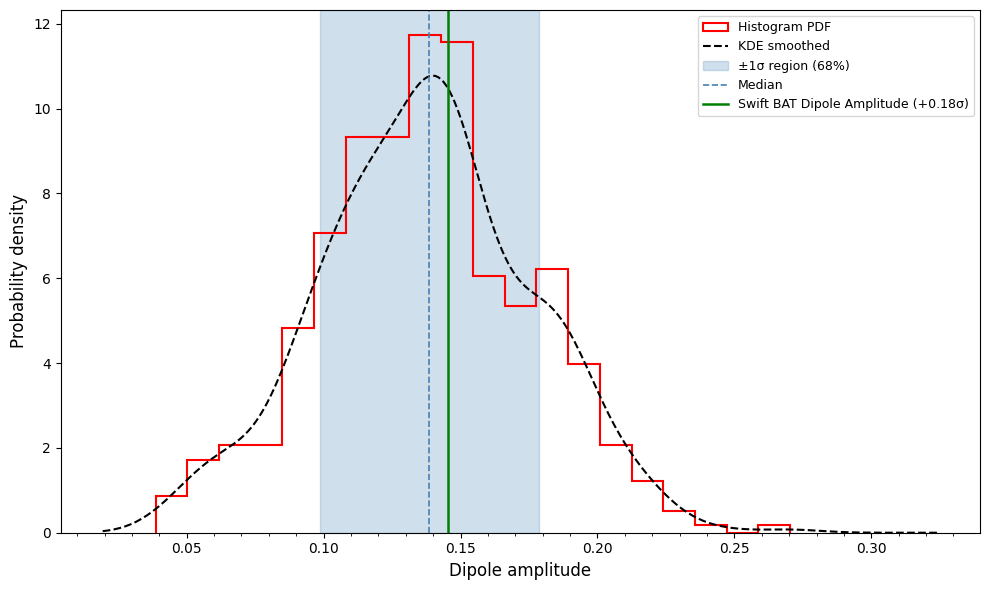}
   
    \caption{Observed dipole for  vs Exposure corrected PDF of dipole amplitude for Swift for $N_{\text{side}}=64$.}
    \label{fig:corrected_dipole_N64}
\end{figure}

 The resulting dipole PDFs of $N_{\text{side}}$ 16, 32, and 64 can be found in  Figs.~\ref{fig:corrected_dipole_N16},~\ref{fig:corrected_dipole_N32}, and \ref{fig:corrected_dipole_N64} respectively. We observe that in all three iterations the observed dipole amplitude lies within $1\sigma$ value of the median of an isotropic distribution. This shows that the dipole estimate is stable across different resolutions.

\begin{figure}[htbp] 
    \centering

    \includegraphics[width=0.85\linewidth]{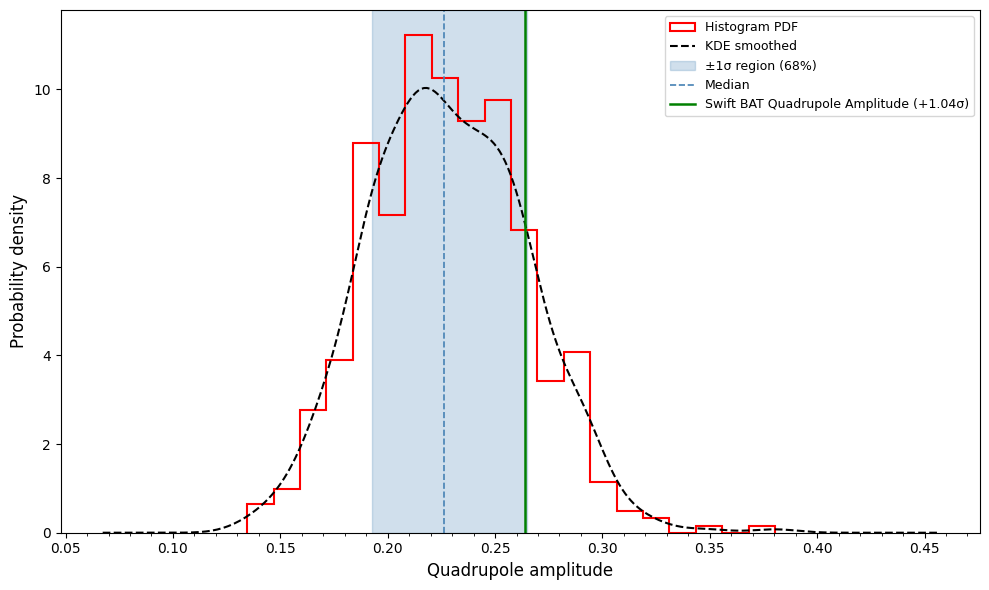}
   
    \caption{Observed quadrupole for  vs Exposure corrected PDF of quadrupole amplitude for Swift for $N_{\text{side}}=16$.}
    \label{fig:corrected_quadrupole_N16}
\end{figure}

\begin{figure}[htbp] 
    \centering

    \includegraphics[width=0.85\linewidth]{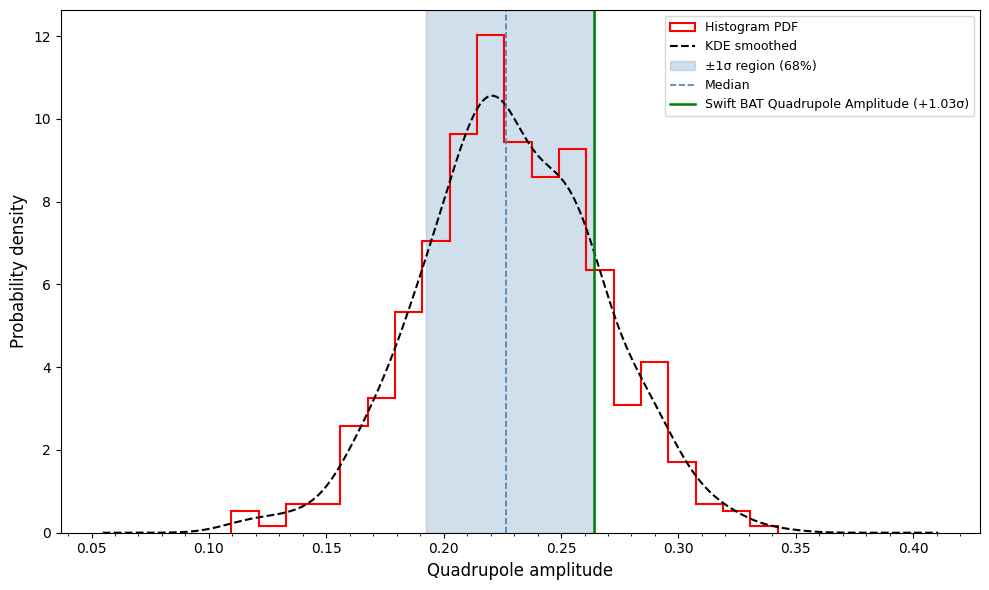}
   
    \caption{Observed quadrupole for  vs Exposure corrected PDF of quadrupole amplitude for Swift for $N_{\text{side}}=32$.}
    \label{fig:corrected_quadrupole_N32}
\end{figure}

\begin{figure}[htbp] 
    \centering

    \includegraphics[width=0.85\linewidth]{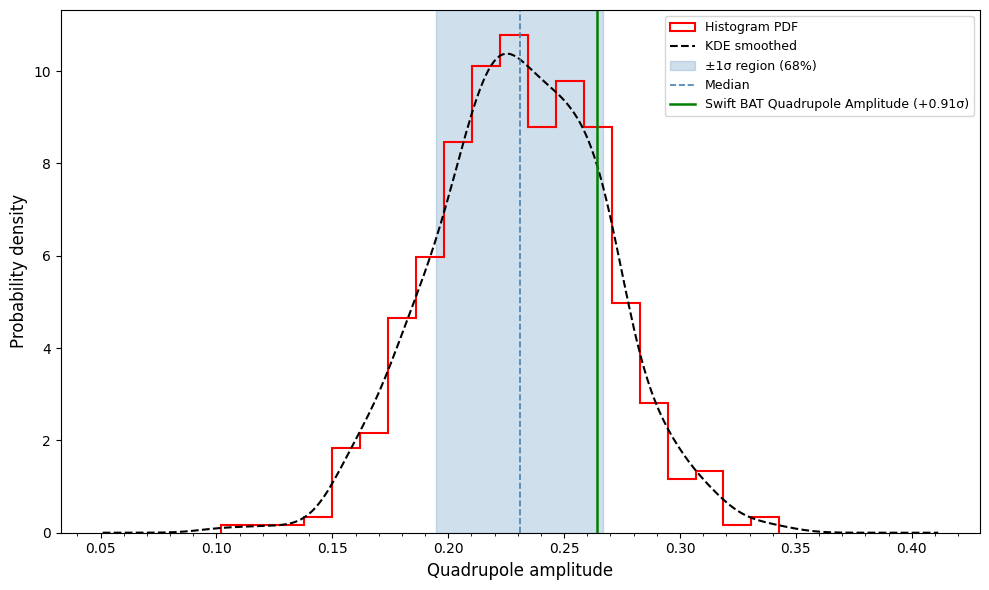}
   
    \caption{Observed quadrupole for  vs Exposure corrected PDF of quadrupole amplitude for Swift for $N_{\text{side}}=64$.}
    \label{fig:corrected_quadrupole_N64}
\end{figure}

A similar result is also observed for quadrupole analyses. Figs.~\ref{fig:corrected_quadrupole_N16},~\ref{fig:corrected_quadrupole_N32}, \ref{fig:corrected_quadrupole_N64} show the resulting quadrupole PDFs of $N_{\text{side}}$ 16, 32, and 64 respectively. The quadrupole values for different resolutions hover around the value of $1\sigma$, thus confirming that our  results are consistent with respect the choice of resolution. A tabular summary of our results can be found in Table~\ref{tab:amplitude_significance}.

\begin{table}[htbp]
    \centering
    \caption{Summary of Swift BAT Dipole and Quadrupole Amplitude Significances}
    \label{tab:amplitude_significance}
    \begin{tabular}{ccc}
        \hline
        \textbf{N} & \textbf{Dipole Amplitude ($\sigma$)} & \textbf{Quadrupole Amplitude ($\sigma$)} \\
        \hline
        16 & +0.3 & +1.0 \\
        32 & +0.3 & +1.0 \\
        64 & +0.2 & +0.9 \\
        \hline
    \end{tabular}
\end{table}

\clearpage

\section{Testing for the split between Short and Long GRBs}
GRBs have been broadly  classified into two categories (short and long) based on their duration ($T_{90}$),  the time over which 90\% of the counts are received~\cite{kouveliotou1993identification}. Long GRBs are characterized by $T_{90} > 2s$ and are attributed to core-collapse of massive stars, while short GRBs are characterized by $T_{90} < 2s$ and are associated with compact binary mergers. Furthermore, some early studies had also found evidence for a $2\sigma$ deviation from isotropy for only the short GRB population~\cite{Maglio03}.  Therefore, it is essential to do the isotropy test separately for the two GRB classes.

For this study we split our initial dataset based on the T90  column in the dataset. We omit  the GRBs for which there is no data on T90. After this split, the dataset consists  of 1483 long GRBs and 144 short GRBs. To test the isotropy, we repeat the entire process of dipole and quadrupole analyses using exposure convolved isotropic skies separately for long and short GRBs. The results are summarized in Table~\ref{tab:grb_amplitudes}.

\begin{figure}[htbp] 
    \centering

    \includegraphics[width=0.85\linewidth]{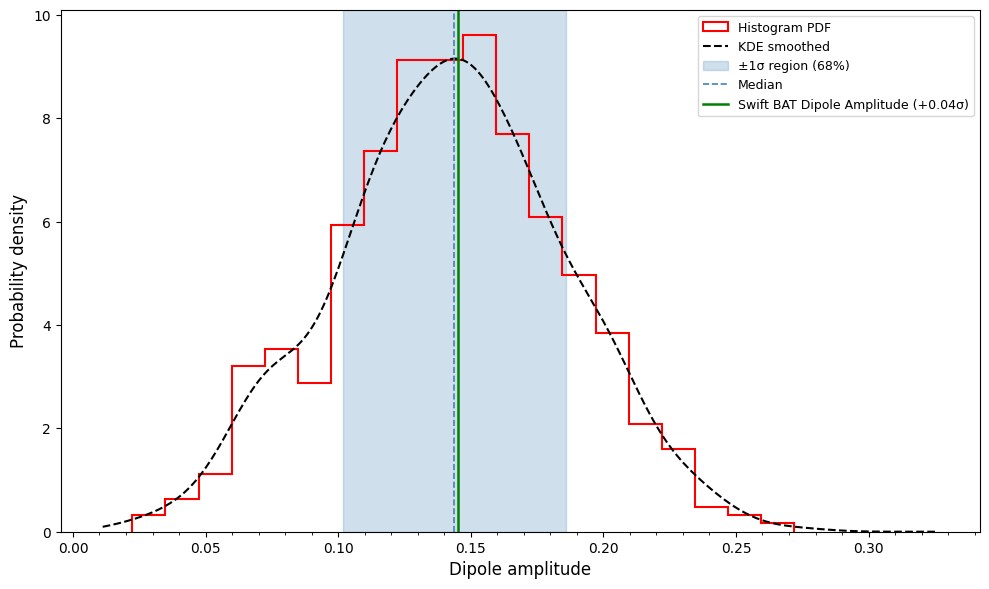}
   
    \caption{Observed dipole for   Exposure corrected PDF of dipole amplitude for long GRBs in Swift. }
    \label{fig:LGRB_corrected_dipole}
\end{figure}

\begin{figure}[htbp] 
    \centering

    \includegraphics[width=0.85\linewidth]{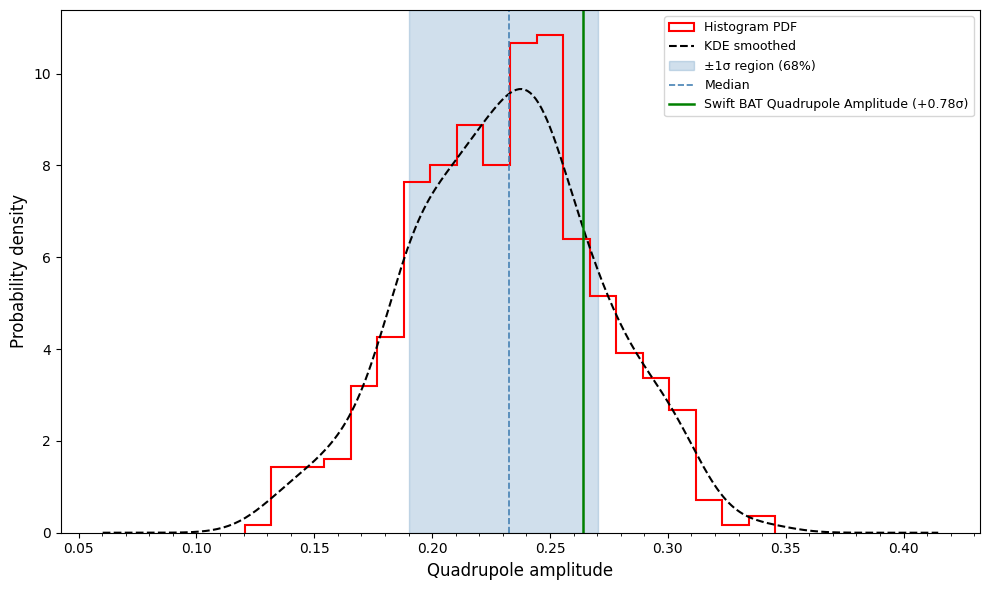}   
    \caption{Observed Quadrupole for   Exposure corrected PDF of dipole amplitude for long GRBs in Swift. }
    \label{fig:LGRB_corrected_quadrupole}
\end{figure}

Figs.~\ref{fig:LGRB_corrected_dipole} and \ref{fig:LGRB_corrected_quadrupole} show the resulting dipole and quadrupole PDFs, respectively, for the long GRB dataset. We find that  the observed dipole and quadrupole amplitude are  within $1\sigma$  of the median of an isotropic distribution. A tabular summary of our results are collated  in Table~\ref{tab:grb_amplitudes}.
This shows that long GRBs follow an isotropic distribution consistent with cosmological principle.

\begin{figure}[htbp] 
    \centering

    \includegraphics[width=0.85\linewidth]{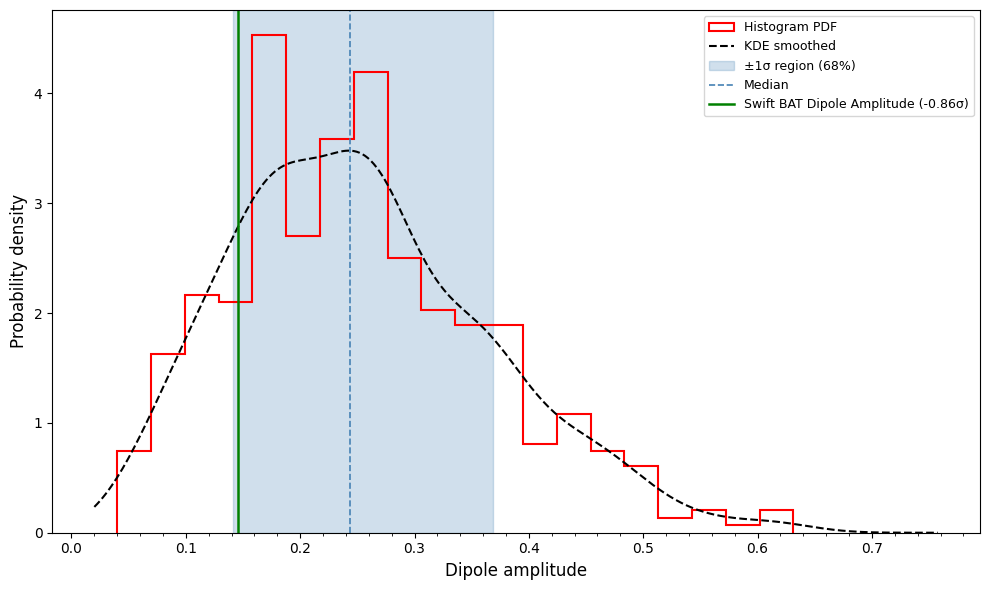}
   
    \caption{Observed dipole for  Exposure corrected PDF of dipole amplitude for short GRBs in Swift.}
    \label{fig:SGRB_corrected_dipole}
\end{figure}

\begin{figure}[htbp] 
    \centering

    \includegraphics[width=0.85\linewidth]{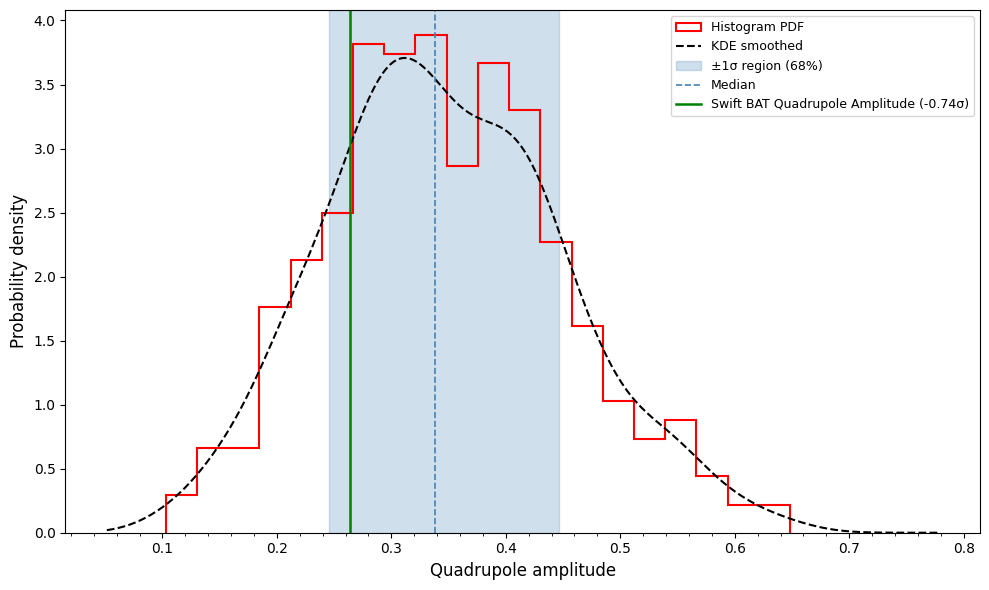}
   
    \caption{Observed Quadrupole for  vs Exposure corrected PDF of dipole amplitude for short GRBs in Swift. }
    \label{fig:SGRB_corrected_quadrupole}
\end{figure}
Similarly, Fig.~\ref{fig:SGRB_corrected_dipole} and \ref{fig:SGRB_corrected_quadrupole} show the resulting dipole and quadrupole PDFs respectively for the short GRB population. Here also, we find the observed dipole and quadrupole amplitude are within $1\sigma$ value of the median of an isotropic distribution. This shows that short GRBs follow an isotropic distribution consistent with cosmological principle.

The independent dipole and quadrupole analyses of the split dataset between long and short GRB shows that they are consistent with isotropy.

\begin{table}[htbp]
\centering
\caption{Comparison of Dipole and Quadrupole Amplitudes for long GRBs and short GRBs relative to the Swift BAT distribution.}
\label{tab:grb_amplitudes}
\begin{tabular}{llccc}
\hline \hline
\textbf{GRB Type} & \textbf{Multipole Component} & \textbf{Median} & \textbf{Swift BAT Amplitude} & \textbf{$\sigma$ Deviation} \\ \hline
Long GRB    & Dipole                       & $\sim$0.144     & $\sim$0.145                  & $+0.04\sigma$               \\
                  & Quadrupole                   & $\sim$0.232     & $\sim$0.264                  & $+0.8\sigma$               \\ \hline
Short GRB   & Dipole                       & $\sim$0.244     & $\sim$0.146                  & $-0.9\sigma$               \\
                  & Quadrupole                   & $\sim$0.339     & $\sim$0.263                  & $-0.7\sigma$               \\ \hline \hline
\end{tabular}
\end{table}
\clearpage

\begin{adjustwidth}{-\extralength}{0cm}
\reftitle{References}



\bibliography{Definitions/references}

%


\end{adjustwidth}
\end{document}